\documentclass[final,1p,times]{elsarticle}

\usepackage{amsthm,amsmath,amsfonts,amssymb}
\usepackage[colorlinks,citecolor=blue,urlcolor=blue,hypertexnames=false]{hyperref}
\usepackage{graphicx}
\usepackage{booktabs}
\usepackage{float}
\usepackage{algorithm}
\usepackage{algorithmicx}
\usepackage{algpseudocode}
\usepackage{enumitem}
\setlist[itemize]{leftmargin=1.4em,itemsep=0.15em,topsep=0.25em}

\journal{Journal of Clinical Epidemiology}
\biboptions{numbers,sort&compress}

\theoremstyle{plain}

\theoremstyle{definition}

\begin{document}

\begin{frontmatter}
\title{Structured Transfer Learning for Survival Risk Stratification in Data-Sparse Clinical Cohorts}

\author[inst1]{Junhan Yu}
\ead{junhan.yu@u.nus.edu}
\author[inst2,inst3,inst4]{Yurui Chen}
\ead{yurui.chen@yale.edu}
\author[inst5]{Juan Delgado-SanMartin}
\ead{j.delgado-san-martin@imperial.ac.uk}
\author[inst2,inst3,inst5]{Dennis Wang\corref{cor1}}
\ead{Dennis\_Wang@a-star.edu.sg}
\author[inst2,inst3]{Hong Pan\corref{cor1}}
\ead{pan\_hong@a-star.edu.sg}
\author[inst1]{Doudou Zhou\corref{cor1}}
\ead{ddzhou@nus.edu.sg}

\cortext[cor1]{Corresponding authors.}

\affiliation[inst1]{
  organization={Department of Statistics \& Data Science, National University of Singapore},
  country={Singapore}}
\affiliation[inst2]{
  organization={Institute for Human Development and Potential, Agency for Science, Technology and Research (A*STAR)},
  country={Singapore}}
\affiliation[inst3]{
  organization={Bioinformatics Institute, Agency for Science, Technology and Research (A*STAR)},
  country={Singapore}}
\affiliation[inst4]{
  organization={Department of Biomedical Informatics \& Data Science, Yale School of Medicine},
  city={New Haven},
  state={CT},
  country={United States}}
\affiliation[inst5]{
  organization={National Heart \& Lung Institute, Imperial College London},
  city={London},
  country={United Kingdom}}

\begin{abstract}
\noindent\textbf{Background:} Survival prediction models are often less reliable when applied to clinical groups with limited sample sizes or few observed outcome events. Models fitted only within these groups may be unstable, whereas models borrowed from larger cohorts may transfer poorly when risk-factor effects differ across populations. We evaluated whether structured transfer learning can improve survival risk stratification in data-sparse clinical cohorts while allowing cohort-specific adaptation.

\par\smallskip\noindent\textbf{Methods:} We developed and evaluated the COhort-shared Rank-rEduced Cox model (CORE-Cox), a two-stage framework for multi-outcome survival prediction. CORE-Cox first learns shared risk-factor patterns across related outcomes in a larger reference cohort using a low-rank Cox coefficient structure, and then adapts these patterns to a smaller cohort of interest through a regularized residual correction. We evaluated CORE-Cox in UK Biobank (White/European source, $n=150{,}093$; Asian target, $n=2{,}534$) and MIMIC-IV (White ICU source, $n=15{,}997$; Asian ICU target, $n=672$), using shared baseline predictors across source and target cohorts. Across nine outcomes in each dataset, CORE-Cox was compared with Cox models fitted only within the smaller cohort, penalized Cox models, low-rank multi-task models, naive pooling, direct model transfer, and single-outcome residual transfer under repeated nested held-out cross-validation.

\par\smallskip\noindent\textbf{Results:} CORE-Cox achieved the best or near-best held-out discrimination across most outcomes. In UK Biobank, the mean concordance index in the smaller cohort improved from 0.733 with standard Cox to 0.766 with CORE-Cox. In MIMIC-IV, the corresponding mean concordance index improved from 0.628 to 0.658, with gains in eight of nine outcomes. CORE-Cox also improved enrichment of individuals ranked in the top 15\% of predicted risk. Hazard-ratio comparisons in selected outcomes showed regularized patterns, with CORE-Cox estimates often intermediate between source-only and target-only models.

\par\smallskip\noindent\textbf{Discussion:} CORE-Cox provides an interpretable structured transfer-learning approach for survival risk stratification in data-sparse clinical cohorts. By combining shared cross-outcome risk structure with cohort-specific residual adaptation, it offers a practical alternative to unstable within-cohort modeling, naive pooling, and unmodified transfer from larger cohorts. Further evaluation is needed before CORE-Cox is used for calibrated absolute-risk prediction or clinical decision-making.
\end{abstract}


\end{frontmatter}

\noindent\textbf{What is new?}

\noindent\textbf{Key findings}
\begin{itemize}[nosep,leftmargin=1.2em]
    \item CORE-Cox improved prediction in data-sparse target cohorts.
    \item Naive pooling and direct transfer were inconsistently helpful.
\end{itemize}

\noindent\textbf{What this adds to what was known?}
\begin{itemize}[nosep,leftmargin=1.2em]
    \item Low-rank source learning enabled cross-outcome borrowing.
    \item Residual adaptation allowed target-specific risk effects.
\end{itemize}

\noindent\textbf{What is the implication and what should change now?}
\begin{itemize}[nosep,leftmargin=1.2em]
    \item Transferability should be tested, not assumed, in small cohorts.
\end{itemize}

\smallskip

\section{Introduction}

Large cohort and electronic health record datasets are increasingly used to develop survival prediction models~\citep{Sudlow2015,https://doi.org/10.13026/hxp0-hg59,Nagai2017-ne}, but the information available for some target populations remains limited. Even when the overall dataset is large, underrepresented ethnic groups, uncommon clinical subgroups, and local care populations may contribute relatively few outcome events. This creates a practical problem for clinical epidemiology: target-only Cox models may be unstable, whereas models borrowed from larger cohorts may not be transportable. Asian participants and patients are included in many large health datasets, but often form smaller analytic cohorts than White or European-background groups, creating a practical setting in which target-cohort survival prediction may be unstable.

This data imbalance is particularly challenging for survival prediction. In time-to-event analyses, right censoring reduces the amount of information available for estimating covariate effects. As a result, Cox models fitted only within these smaller groups can produce unstable risk scores and imprecise hazard-ratio estimates~\citep{Cox1972}. These limitations are important in clinical epidemiology because risk stratification models are often expected to inform prevention, monitoring, or follow-up strategies in groups for which reliable model development data may be limited.

Borrowing information from larger, better-represented cohorts is therefore appealing, but it is not automatically safe. Directly applying a model trained in a larger cohort, or naively pooling larger and smaller cohorts, assumes that risk-factor effects are sufficiently transportable across populations. This assumption may fail when cohorts differ in demographics, case mix, measurement processes, healthcare access, or clinical practice. Such differences can induce covariate shift and effect heterogeneity, which may reduce model transportability and raise concerns about equitable risk stratification if models are applied without target-cohort evaluation~\citep{Obermeyer2019,Vyas2020,pmlr-v125-jeong20a,Qiu2023}. When the larger cohort dominates the estimation process, pooled models may also obscure associations that are specific to the smaller target cohort.

A second opportunity arises because many clinically important time-to-event outcomes are related. Cardiometabolic and chronic disease outcomes share risk factors and biological pathways, including inflammation, metabolic dysfunction, and vascular injury~\citep{James2021,Zhao2017,Lu2024,Furman2019,Bennett2018,DelGiudice2018}. Multi-task learning can use this cross-outcome relatedness to improve statistical efficiency when event counts are limited~\citep{Caruana1997-es}. Low-rank modeling provides one way to represent multiple outcome-specific coefficient vectors through a smaller number of shared latent risk patterns~\citep{Yuan2007,Zhang2023}. However, existing multi-task survival methods are usually designed for a single data environment, whereas many survival transfer-learning methods adapt one outcome at a time and do not fully use cross-outcome structure~\citep{Li2016,Li02102023}.

We developed the COhort-shared Rank-rEduced Cox model (CORE-Cox) to address this gap. CORE-Cox first estimates a low-rank Cox coefficient matrix across related outcomes in a larger source cohort, then adapts this structure to a smaller target cohort through a regularized residual correction. This two-stage framework is designed to borrow shared risk information while allowing target-specific departures from the source model. It preserves the interpretability of Cox regression and represents cross-outcome relatedness through coefficient structure rather than through a joint distribution of multiple event times.

In this study, we evaluated CORE-Cox in two health-data settings: UK Biobank and MIMIC-IV. In both settings, Asian cohorts were used as target cohorts of interest, and the corresponding larger White or European-background cohorts were used as source cohorts. We compared CORE-Cox with target-only Cox models, penalized Cox models, low-rank multi-task models, naive pooling, direct source transfer, and single-outcome residual transfer. The evaluation focused on held-out discrimination using the concordance index, high-risk identification using top-15\% event-rate lift, and stability of hazard-ratio estimates. Our aim was to assess whether structured information borrowing can improve survival risk stratification in data-sparse target cohorts while reducing the risk of negative transfer.
\section{Methods}
\label{sec:methods}

\subsection{Study Design, Reporting Guidance, and Data Sources}

We conducted a model-development study with repeated nested internal validation to evaluate structured transfer learning for time-to-event prediction in data-sparse target cohorts. The analysis used two health-data settings: UK Biobank, a population-based prospective cohort, and MIMIC-IV, an intensive-care electronic health record (EHR) database~\citep{Sudlow2015,https://doi.org/10.13026/hxp0-hg59}. The study was reported in accordance with TRIPOD+AI for clinical prediction model studies~\citep{Collins2024TRIPODAI}. Because the empirical evaluations used observational cohort and EHR data sources, STROBE was also used where applicable to report cohort selection, data sources, outcome definitions, follow-up, missing data, and validation procedures~\citep{vonElm2007STROBE}. Completed TRIPOD+AI, TRIPOD+AI for Abstracts, and STROBE checklists are provided as Supplementary Material. In each setting, a larger source cohort was used to learn shared survival risk structure across related outcomes, and a smaller target cohort was used for target-specific model adaptation and held-out evaluation. We also conducted supplementary simulation analyses under known coefficient structures to evaluate coefficient-estimation accuracy. Detailed data extraction, sample selection, variable definitions, preprocessing procedures, and simulation settings are provided in the Supplementary Material.

\subsection{Source and Target Cohorts}

In the UK Biobank analysis, cohort membership was defined using self-reported ethnicity. The source cohort included participants who self-reported White or related European ethnic backgrounds, and the target cohort included participants who self-reported Asian ethnic backgrounds. The final analytic sample included 150,093 source-cohort participants and 2,534 target-cohort participants.

In the MIMIC-IV analysis, cohort membership used race or ethnicity categories recorded in the EHR. The source cohort consisted of 15,997 ICU patients recorded as White, and the target cohort consisted of 672 ICU patients recorded as Asian. In both datasets, participants or patients who self-reported or were recorded as Asian were used as the target group for model adaptation and held-out evaluation.

\subsection{Outcomes, Follow-Up, and Predictors}

Each dataset included nine outcome-specific time-to-event prediction tasks. In UK Biobank, the outcomes were type 2 diabetes, hypertension, coronary artery disease, stroke, heart failure, non-alcoholic fatty liver disease, multimorbidity, all-cause mortality, and death by multimorbidity. In MIMIC-IV, the outcomes were type 2 diabetes, hypertension, coronary artery disease, stroke, heart failure, hypertensive chronic kidney disease, liver disease, multimorbidity, and death.

For UK Biobank, baseline was defined as March 11, 2011, the mean administration date of the dietary questionnaire used in this analysis. This date was used to align follow-up with the availability of diet-related baseline predictors collected through the 2011 dietary questionnaire window, and follow-up time was defined from baseline to the first recorded incident event or censoring time. For MIMIC-IV, baseline was defined as the time of initial ICU admission. MIMIC-IV outcome status was derived from diagnosis codes recorded during the current hospitalization, including discharge diagnoses; these outcomes should therefore be interpreted as recorded hospitalization diagnoses rather than confirmed incident disease onset after ICU admission. Time-to-event variables in MIMIC-IV used the available recorded diagnosis or hospitalization timing relative to ICU admission, and patients without an observed outcome were treated as right-censored. The same individual could contribute to more than one outcome-specific risk set, with event time, censoring, and risk-set membership defined separately for each outcome.

Predictors were selected from baseline demographic, lifestyle, socioeconomic, clinical, laboratory, or EHR-derived variables according to the data source. CORE-Cox requires a shared covariate representation between the source and target cohorts within each dataset. Complete-case analysis was used to preserve a common source-target covariate representation for each data source; no missing-value imputation was performed. Dataset-specific covariate definitions, preprocessing details, complete-case sample construction, and outcome event counts for source and target cohorts are reported in the Supplementary Material.

\subsection{CORE-Cox Model Overview}

CORE-Cox is a two-stage extension of Cox regression designed to borrow information from a larger source cohort while allowing adaptation to a smaller target cohort. For a standard Cox model, the hazard for an individual with covariate vector $\mathbf{x}$ is
\[
h(t|\mathbf{x}) = h_0(t)\exp(\mathbf{x}^{\top}\boldsymbol{\beta}),
\]
where $h_0(t)$ is an unspecified baseline hazard and $\boldsymbol{\beta}$ is the vector of log-hazard ratios.

For $K$ related outcome-specific prediction tasks and $p$ shared predictors, CORE-Cox collects the outcome-specific Cox coefficient vectors into a coefficient matrix $\mathbf{B}\in\mathbb{R}^{p\times K}$. Each column of $\mathbf{B}$ corresponds to one outcome-specific Cox model, and each row corresponds to one predictor across outcomes.

In the first stage, CORE-Cox estimates shared risk-factor structure in the source cohort by imposing a low-rank representation on the source coefficient matrix:
\[
\mathbf{B}^{(s)} = \mathbf{U}^{(s)}(\mathbf{V}^{(s)})^{\top}.
\]
The rank controls the number of latent coefficient patterns shared across outcomes. This stage allows related outcomes to share information while preserving outcome-specific Cox coefficient vectors.

In the second stage, CORE-Cox adapts the source-derived coefficient matrix to the target cohort through a residual correction:
\[
\mathbf{B}^{(t)} = \hat{\mathbf{B}}^{(s)} + \mathbf{\Theta}.
\]
Here, $\hat{\mathbf{B}}^{(s)}$ is the estimated source coefficient matrix and $\mathbf{\Theta}$ captures target-specific departures from the source model. The residual correction is estimated in the target cohort using a penalized Cox partial-likelihood objective. The residual penalty controls the degree of borrowing: stronger penalization shrinks the target model toward the source-derived structure, whereas weaker penalization allows greater target-specific adaptation.

Operationally, Stage 1 uses only the larger source cohort to jointly fit the outcome-specific Cox models under the low-rank constraint. In Stage 2, the estimated source coefficient matrix is held fixed, and only target-cohort data are used to estimate the residual matrix. This design allows the target model to borrow source-derived cross-outcome structure while deviating from it when target-cohort event data support different covariate effects. The low-rank dimension and residual penalty were selected within the inner cross-validation procedure described below.

For each target-cohort individual and outcome, CORE-Cox uses the adapted coefficient vector to compute a Cox log-risk score. These scores were used for held-out risk ranking, concordance-index estimation, and top-15\% event-rate lift rather than for calibrated absolute-risk prediction. CORE-Cox therefore shares information across outcomes through the coefficient matrix without requiring a joint model for multiple event times. Detailed objective functions, regularization terms, optimization steps, and baseline-hazard estimation details are provided in the Supplementary Material.

\subsection{Comparator Models}

CORE-Cox was compared with seven benchmark methods: Cox, Cox-Lasso, Cox-Ridge, LR-MTL-Target, LR-MTL-Source, LR-MTL-Both, and Cox-Transfer. These benchmarks represented target-only fitting, penalized target-only fitting, low-rank multi-task learning, direct source transfer, naive source-target pooling, and single-outcome residual transfer.

Briefly, Cox, Cox-Lasso, and Cox-Ridge were fitted separately for each target-cohort outcome. LR-MTL-Target used only target-cohort data with low-rank multi-task sharing across outcomes. LR-MTL-Source used only source-cohort data and was directly applied to the target cohort. LR-MTL-Both pooled source and target individuals before fitting a low-rank multi-task model. Cox-Transfer used single-outcome residual adaptation from source to target. Full benchmark definitions and source/target data use are provided in Supplementary Section~S7.

\subsection{Model Tuning and Held-Out Validation}

Predictive performance was evaluated using repeated nested cross-validation. In each repeat, an outer 5-fold split was used for held-out evaluation. For each outer split, one fold was held out for evaluation and the remaining folds were used for model development. Hyperparameters, including the low-rank dimension and regularization parameters, were selected within the outer training data using inner 4-fold cross-validation.

The held-out outer fold was not used for hyperparameter tuning, model selection, or outcome-informed preprocessing. After tuning, each model was refitted on the full outer training set and evaluated on the held-out fold. The same outer evaluation splits were used for CORE-Cox and all benchmark models, and real-data analyses were repeated across 100 random seeds.

For real-data prediction analyses, hyperparameters were selected by maximizing the mean validation C-index across outcome tasks in the inner cross-validation. For these supplementary simulation analyses, coefficient-estimation performance was assessed using coefficient error metrics because the true coefficient matrix was known.

\subsection{Performance Measures}

The primary real-data evaluation focused on held-out risk ranking and high-risk identification in the target cohorts. Discrimination was assessed using Harrell's concordance index computed on held-out outer folds; censoring was handled by the concordance estimator through comparable pairs, and no inverse-probability-of-censoring weighting was applied. High-risk identification was assessed using top-15\% event-rate lift. Within each held-out fold and outcome, individuals were ranked by predicted log-risk, the top 15\% were selected, and the observed event proportion in this high-risk group was divided by the observed event proportion in the full held-out fold. The 15\% threshold was used as a pragmatic high-risk stratum that retained enough events for stable estimation while remaining selective for risk-enrichment evaluation. Estimate stability was examined in representative analyses by comparing hazard-ratio estimates and 95\% confidence intervals across methods.

\begin{figure}[h!]
    \centering
    \includegraphics[width=0.9\linewidth]{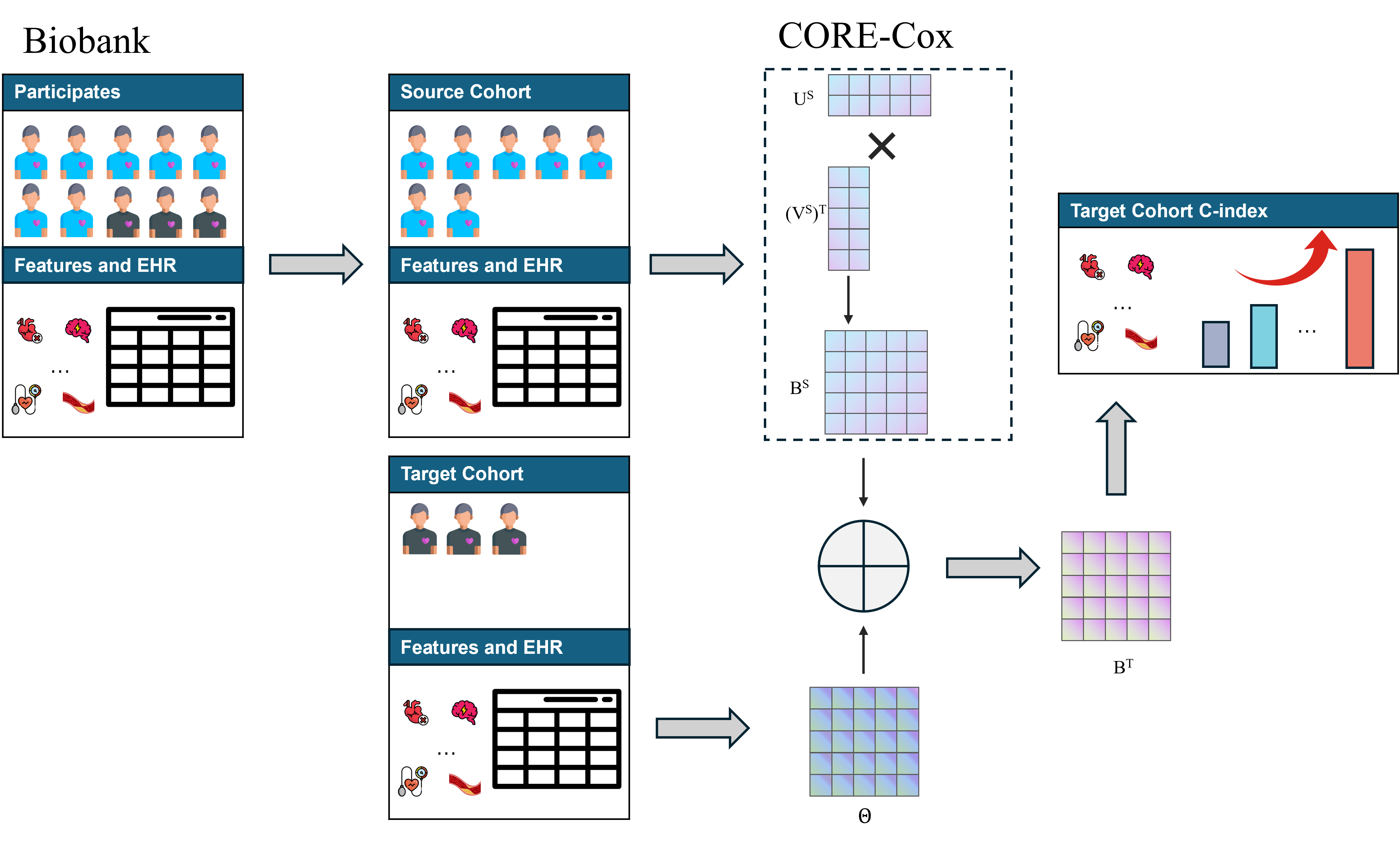}
   \caption{Schematic overview of CORE-Cox. Stage~1 learns shared cross-outcome Cox coefficient patterns in the source cohort. Stage~2 adapts the source-derived coefficients to the target cohort through a penalized residual correction.}
    \label{fig:framework}
\end{figure}

\section{Results}

\subsection{UK Biobank Target-Cohort Prediction}
\label{sec:ukb_data}

In UK Biobank, target-cohort prediction was evaluated across nine chronic disease outcomes in 2,534 Asian participants, using 150,093 White/European participants as the source cohort.

\begin{figure}[h!]
    \centering
    \includegraphics[width=1\linewidth]{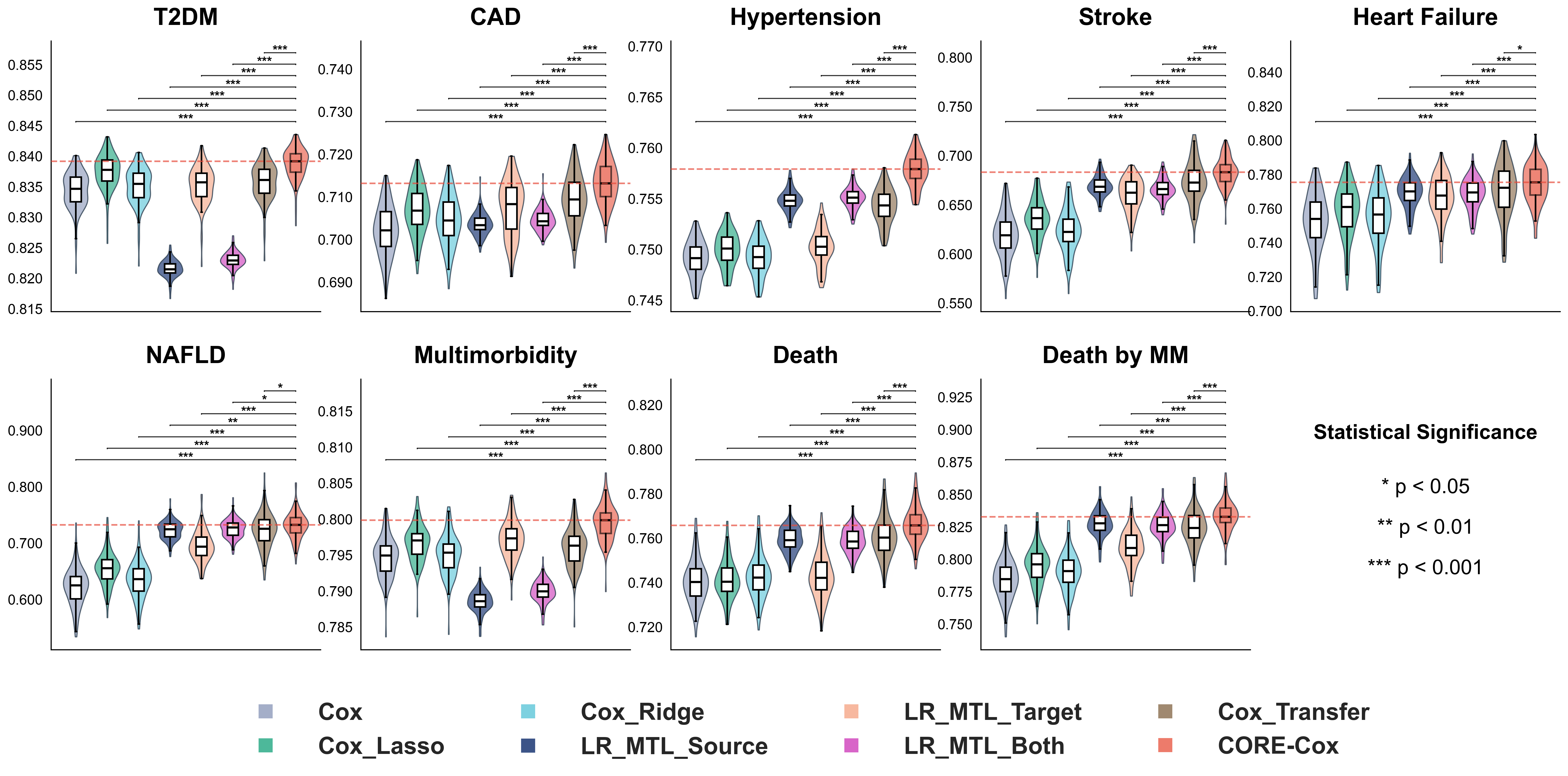}
    \caption{C-index comparison across nine outcomes in the UK Biobank Asian cohort. Violins summarize repeated evaluations by method and outcome.}
    \label{fig:real_data_results}
\end{figure}

Across repeated held-out validation, CORE-Cox achieved the highest average C-index in the Asian target cohort, with a mean C-index of 0.766 compared with 0.733 for standard target-only Cox. CORE-Cox had the highest mean C-index for each of the nine outcomes, with particularly large gains for NAFLD and stroke. Full outcome-specific C-index results are reported in Table~S5 and summarized in Figure~\ref{fig:real_data_results}.

As a clinical benchmark for the magnitude of these gains, we compared the improvement from CORE-Cox with the discrimination gain obtained by adding age to baseline models; details are provided in the Supplementary Material.

\subsection{MIMIC-IV ICU Target-Cohort Prediction}

We next evaluated CORE-Cox in MIMIC-IV, an ICU EHR setting, using 15,997 ICU patients recorded as White as the source cohort and 672 ICU patients recorded as Asian as the target cohort. The benchmark definitions, tuning strategy, and held-out evaluation protocol were the same as in the UK Biobank analysis.

\begin{figure}[h!]
    \centering
    \includegraphics[width=1\linewidth]{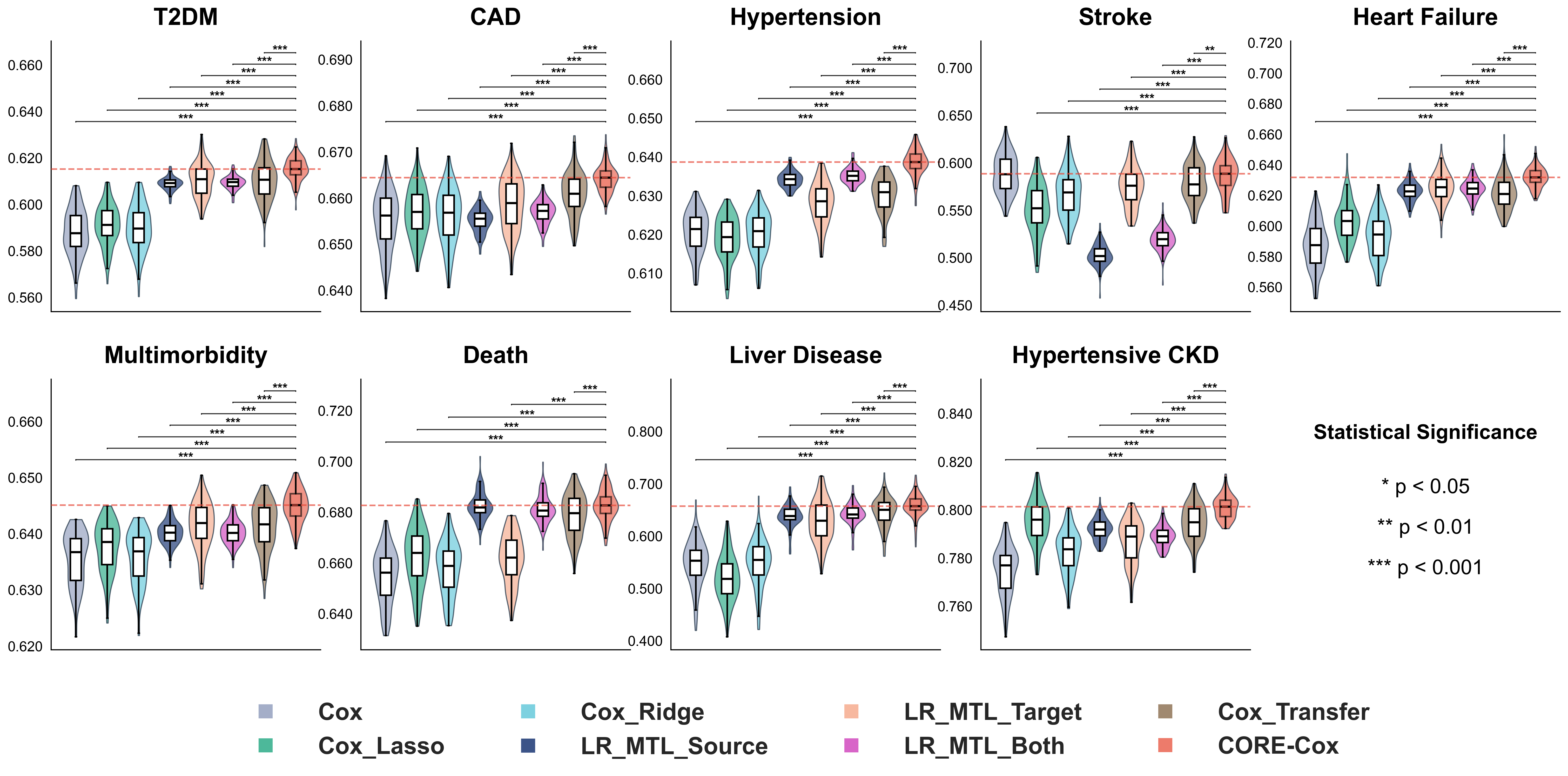}
    \caption{C-index comparison across nine outcomes in the MIMIC-IV Asian ICU cohort. Violins summarize repeated evaluations by method and outcome.}
    \label{fig:mimic_results_c_index}
\end{figure}

CORE-Cox improved discrimination over standard target-only Cox for eight of the nine outcomes. The mean C-index was 0.658 for CORE-Cox and 0.628 for standard Cox. The main exception was stroke, where CORE-Cox did not outperform target-only Cox, possibly reflecting limited target-cohort event counts and outcome mechanisms tied to acute ICU physiology. Full outcome-specific results are reported in Table~S7 and summarized in Figure~\ref{fig:mimic_results_c_index}.

\subsection{High-Risk Identification}

CORE-Cox also improved identification of individuals or patients at elevated predicted risk. Among those ranked in the top 15\% of predicted risk, the average event-rate lift was 3.16 across UK Biobank outcomes and 1.85 across MIMIC-IV outcomes (Tables~S6 and S8). Gains were largest in settings where naive pooling or direct source transfer degraded relative to target-only fitting, including type 2 diabetes and multimorbidity in UK Biobank and hypertensive chronic kidney disease in MIMIC-IV.

\subsection{Transfer Behavior and Estimation Stability}

Hazard-ratio estimates supported the interpretation that CORE-Cox improved coefficient stability by combining source-cohort information with target-specific adaptation. For NAFLD in the UK Biobank Asian cohort, CORE-Cox produced narrower confidence intervals than target-only Cox, and estimates often fell between source-only and target-only estimates (Figure~\ref{fig:nafld_forest_plot}). Similar patterns are shown in Supplementary Section~S5.

\begin{figure}[h!]
    \centering
    \includegraphics[width=1\linewidth]{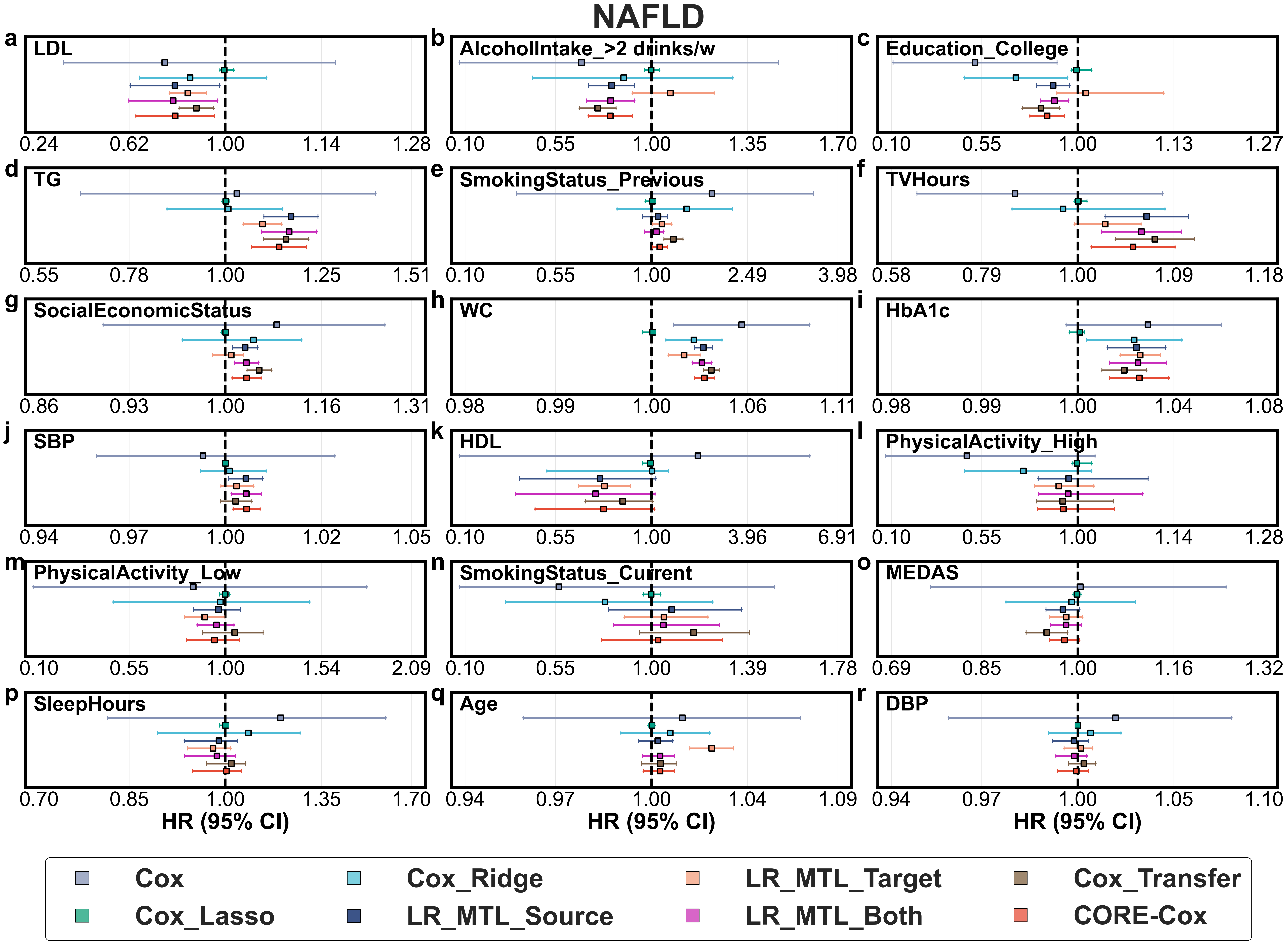}
    \caption{Hazard-ratio estimates and 95\% confidence intervals for selected NAFLD risk factors in the UK Biobank Asian cohort.}
    \label{fig:nafld_forest_plot}
\end{figure}

These findings suggest that CORE-Cox did not simply reproduce the source model or rely only on sparse target-cohort data; instead, it adapted source-derived structure toward target-cohort associations.

\subsection{Simulation Analysis of Estimation Stability}
\label{sec:simulation}

Simulation experiments were used to assess coefficient-estimation accuracy under controlled source-target heterogeneity. These analyses evaluated whether structured borrowing could reduce estimation error in small target cohorts while avoiding the limitations of unmodified source transfer or naive source-target pooling.

Across simulated settings, CORE-Cox reduced coefficient-estimation error relative to target-only Cox, with the largest gains when the target sample size was small and source-target differences were moderate. In the bootstrap coverage analysis, CORE-Cox achieved empirical 95\% interval coverage of 0.929 with an average interval width of 0.273, compared with coverage of 0.941 and average width of 0.477 for target-only Cox. Thus, CORE-Cox produced substantially narrower intervals while maintaining comparable empirical coverage. Full data-generating settings, RRMSE curves, tuning details, and coverage results are provided in the Supplementary Material.

\section{Discussion}

This study evaluated CORE-Cox as a structured transfer-learning approach for survival risk stratification when the target clinical cohort is much smaller than a related source cohort. Across a population biobank setting and an ICU EHR setting, CORE-Cox improved held-out discrimination and high-risk identification relative to standard target-only Cox models, suggesting that structured information borrowing can be useful beyond a single cohort type.

The main methodological contribution is the separation of shared risk structure from target-specific adaptation. Rather than pooling heterogeneous cohorts or transferring a source model without modification, CORE-Cox learns a low-rank multi-outcome Cox coefficient matrix in the source cohort and then estimates a residual correction in the target cohort. This lets the model borrow strength from larger cohorts while retaining a mechanism for target-cohort differences. Direct transfer and naive pooling were competitive for some outcomes but degraded for others, whereas CORE-Cox more consistently remained among the top-performing methods. This places CORE-Cox between complete pooling, which assumes source and target data are sufficiently exchangeable, and target-only fitting, which ignores potentially useful source information.

Together, these findings support CORE-Cox as a transferable risk-stratification framework for data-sparse target cohorts. The two-stage structure may also be useful in multi-site settings where individual-level source data cannot be shared. A source site could share fitted model outputs, while the target site performs adaptation locally~\citep{Maro2009, Brown2020}. Formal federated implementations were outside the scope of this study.

Several limitations should be considered. First, we used broad ethnicity categories to construct a practical source-target setting with substantial target-cohort data sparsity. These categories are coarse and should not be interpreted as biologically homogeneous groups. Observed source-target differences may reflect not only ethnicity-related factors, but also differences in covariate distributions, baseline risk, health-care access, measurement processes, and other contextual factors. Second, the study was retrospective and used secondary data; therefore, outcome definitions, diagnosis timing, censoring, and missingness may differ across cohorts and data systems. In MIMIC-IV, chronic disease outcomes were based on diagnoses recorded during the current hospitalization, including discharge diagnoses. These outcomes should therefore be interpreted as recorded hospitalization diagnoses rather than adjudicated incident disease onset after ICU admission. More generally, diagnosis time in EHR data may not correspond to true disease onset, and competing events may affect observed outcomes.
Technical limitations include the single-source, single-target design, the use of selected conventional predictors, and the absence of high-dimensional biomarkers or multi-omics data. We also did not develop formal large-sample theory for the two-stage penalized estimator. Future work should address source-model uncertainty, model-selection uncertainty, multiple candidate source cohorts, and data-adaptive weighting to reduce negative transfer from poorly matched sources.

In conclusion, CORE-Cox provides an interpretable framework for transferable survival risk stratification in small target cohorts. By combining low-rank multi-outcome learning with target-specific residual adaptation, it offers a practical alternative to target-only modeling, naive pooling, and unmodified source transfer. Further deployment-oriented validation will be needed to assess calibrated absolute-risk prediction, threshold-based clinical utility, and clinical decision-making applications.
\section*{Supplementary Material}
The supplementary material contains cohort definitions, preprocessing details, extended performance tables and figures, benchmark definitions, validation protocol, technical model specification, and simulation results.

\section*{Ethics Approval and Consent to Participate}
This study used secondary data from UK Biobank and MIMIC-IV. UK Biobank has approval from the North West Multi-centre Research Ethics Committee, and data for this study were accessed under UK Biobank Application ID 523060. MIMIC-IV is a de-identified critical-care database; creation and sharing of the resource were approved by the institutional review boards of Beth Israel Deaconess Medical Center and the Massachusetts Institute of Technology, with a waiver of informed consent. The present analysis involved no new participant recruitment or direct participant contact.

\section*{Study Registration}
This was a methodological model-development and evaluation study rather than an inferential hypothesis-testing observational study. No prospective protocol was registered before analysis. Model specification, benchmark definitions, hyperparameter tuning, validation procedures, and code availability are described in the manuscript and Supplementary Material.

\section*{Patient and Public Involvement}
Patients and members of the public were not involved in the design, conduct, reporting, or dissemination planning for this methodological secondary-data study.

\section*{CRediT Authorship Contribution Statement}
Junhan Yu: Conceptualization, Methodology, Software, Formal analysis, Data curation, Visualization, Writing - original draft. Yurui Chen: Investigation, Writing - review and editing. Juan Delgado-SanMartin: Investigation, Writing - review and editing. Dennis Wang: Conceptualization, Methodology, Supervision, Writing - review and editing. Hong Pan: Conceptualization, Methodology, Supervision, Writing - review and editing. Doudou Zhou: Conceptualization, Methodology, Supervision, Writing - review and editing.

\section*{Funding}
Dennis Wang is supported by the Academy of Medical Sciences Professorship (APR7.1002), A*STAR Early Childhood Grant (H24P2M0005), and Open Fund-Large Collaborative Grant (OF-LCG; MOH-000504). Hong Pan is supported by the Singapore Ministry of Health's National Medical Research Council (NMRC) under the Population Health Research Grant, New Investigator Grant (MOH-001424-00). The funders had no role in study design, data analysis, interpretation of results, manuscript writing, or the decision to submit the article for publication.

\section*{Data Availability}
UK Biobank data were accessed under Application ID 523060 and are available to approved researchers through UK Biobank. MIMIC-IV data are de-identified and available to credentialed users through PhysioNet after completion of the required data-use training. Individual-level data cannot be redistributed by the authors.

\section*{Code Availability}
Analysis code will be provided as supplementary submission material.

\section*{Declaration of Competing Interest}
The authors declare that they have no known competing financial interests or personal relationships that could have appeared to influence the work reported in this paper.

\section*{Supplementary Data}
Supplementary material accompanies this article.
\clearpage

\bibliographystyle{elsarticle-num-names}

\bibliography{ref}

@article{Yuan2007,
  title = {Dimension Reduction and Coefficient Estimation in Multivariate Linear Regression},
  author = {Yuan, Ming and Ekici, Ali and Lu, Zhaosong and Monteiro, Renato},
  journal = {Journal of the Royal Statistical Society: Series B (Statistical Methodology)},
  volume = {69},
  number = {3},
  pages = {329--346},
  year = {2007},
  publisher = {Royal Statistical Society and Wiley},
  url = {http://www.jstor.org/stable/4623272}
}

@inproceedings{Li2016,
  title = {Transfer Learning for Survival Analysis via Efficient L2, 1-Norm Regularized Cox Regression},
  url = {http://dx.doi.org/10.1109/ICDM.2016.0034},
  DOI = {10.1109/icdm.2016.0034},
  booktitle = {2016 IEEE 16th International Conference on Data Mining (ICDM)},
  publisher = {IEEE},
  author = {Li,  Yan and Wang,  Lu and Wang,  Jie and Ye,  Jieping and Reddy,  Chandan K.},
  year = {2016},
  month = dec,
  pages = {231–240}
}

@InProceedings{pmlr-v125-jeong20a,
  title = 	 {Robust causal inference under covariate shift via worst-case subpopulation treatment effects},
  author =       {Jeong, Sookyo and Namkoong, Hongseok},
  booktitle = 	 {Proceedings of Thirty Third Conference on Learning Theory},
  pages = 	 {2079--2084},
  year = 	 {2020},
  editor = 	 {Abernethy, Jacob and Agarwal, Shivani},
  volume = 	 {125},
  series = 	 {Proceedings of Machine Learning Research},
  month = 	 {09--12 Jul},
  publisher =    {PMLR},
  url = 	 {https://proceedings.mlr.press/v125/jeong20a.html}
}

@article{Li02102023,
author = {Ziyi Li and Yu Shen and Jing Ning},
title = {Accommodating Time-Varying Heterogeneity in Risk Estimation under the Cox Model: A Transfer Learning Approach},
journal = {Journal of the American Statistical Association},
volume = {118},
number = {544},
pages = {2276--2287},
year = {2023},
publisher = {Taylor \& Francis},
doi = {10.1080/01621459.2023.2210336},
    note ={PMID: 38505403},
URL = { 
    
        https://doi.org/10.1080/01621459.2023.2210336
},
eprint = { 
    
        https://doi.org/10.1080/01621459.2023.2210336
}
}

@article{Zhang2023,
  title = {Personalized prediction for multiple chronic diseases by developing the multi-task Cox learning model},
  volume = {19},
  ISSN = {1553-7358},
  url = {http://dx.doi.org/10.1371/journal.pcbi.1011396},
  DOI = {10.1371/journal.pcbi.1011396},
  number = {9},
  journal = {PLOS Computational Biology},
  publisher = {Public Library of Science (PLoS)},
  author = {Zhang,  Shuaijie and Yang,  Fan and Wang,  Lijie and Si,  Shucheng and Zhang,  Jianmei and Xue,  Fuzhong},
  editor = {Lofgren,  Eric},
  year = {2023},
  month = sep,
  pages = {e1011396}
}

@article{Maro2009,
  title = {Design of a National Distributed Health Data Network},
  volume = {151},
  ISSN = {1539-3704},
  url = {http://dx.doi.org/10.7326/0003-4819-151-5-200909010-00139},
  DOI = {10.7326/0003-4819-151-5-200909010-00139},
  number = {5},
  journal = {Annals of Internal Medicine},
  publisher = {American College of Physicians},
  author = {Maro,  Judith C. and Platt,  Richard and Holmes,  John H. and Strom,  Brian L. and Hennessy,  Sean and Lazarus,  Ross and Brown,  Jeffrey S.},
  year = {2009},
  month = sep,
  pages = {341–344}
}

@article{Qiu2023,
  title = {Prediction sets adaptive to unknown covariate shift},
  volume = {85},
  ISSN = {1467-9868},
  url = {http://dx.doi.org/10.1093/jrsssb/qkad069},
  DOI = {10.1093/jrsssb/qkad069},
  number = {5},
  journal = {Journal of the Royal Statistical Society Series B: Statistical Methodology},
  publisher = {Oxford University Press (OUP)},
  author = {Qiu,  Hongxiang and Dobriban,  Edgar and Tchetgen Tchetgen,  Eric},
  year = {2023},
  month = jul,
  pages = {1680–1705}
}

@article{DelGiudice2018,
  title = {Rethinking IL-6 and CRP: Why they are more than inflammatory biomarkers,  and why it matters},
  volume = {70},
  ISSN = {0889-1591},
  url = {http://dx.doi.org/10.1016/j.bbi.2018.02.013},
  DOI = {10.1016/j.bbi.2018.02.013},
  journal = {Brain,  Behavior,  and Immunity},
  publisher = {Elsevier BV},
  author = {Del Giudice,  Marco and Gangestad,  Steven W.},
  year = {2018},
  month = may,
  pages = {61–75}
}

@article{Lu2024,
  title = {Type 2 diabetes mellitus in adults: pathogenesis,  prevention and therapy},
  volume = {9},
  ISSN = {2059-3635},
  url = {http://dx.doi.org/10.1038/s41392-024-01951-9},
  DOI = {10.1038/s41392-024-01951-9},
  number = {1},
  journal = {Signal Transduction and Targeted Therapy},
  publisher = {Springer Science and Business Media LLC},
  author = {Lu,  Xi and Xie,  Qingxing and Pan,  Xiaohui and Zhang,  Ruining and Zhang,  Xinyi and Peng,  Ge and Zhang,  Yuwei and Shen,  Sumin and Tong,  Nanwei},
  year = {2024},
  month = oct 
}

@article{Zhao2017,
  title = {Identification of new susceptibility loci for type 2 diabetes and shared etiological pathways with coronary heart disease},
  volume = {49},
  ISSN = {1546-1718},
  url = {http://dx.doi.org/10.1038/ng.3943},
  DOI = {10.1038/ng.3943},
  number = {10},
  journal = {Nature Genetics},
  publisher = {Springer Science and Business Media LLC},
  author = {Zhao,  Wei and Rasheed,  Asif and Tikkanen,  Emmi and Lee,  Jung-Jin and Butterworth,  Adam S and Howson,  Joanna M M and Assimes,  Themistocles L and Chowdhury,  Rajiv and Orho-Melander,  Marju and Damrauer,  Scott and Small,  Aeron and Asma,  Senay and Imamura,  Minako and Yamauch,  Toshimasa and Chambers,  John C and Chen,  Peng and Sapkota,  Bishwa R and Shah,  Nabi and Jabeen,  Sehrish and Surendran,  Praveen and Lu,  Yingchang and Zhang,  Weihua and Imran,  Atif and Abbas,  Shahid and Majeed,  Faisal and Trindade,  Kevin and Qamar,  Nadeem and Mallick,  Nadeem Hayyat and Yaqoob,  Zia and Saghir,  Tahir and Rizvi,  Syed Nadeem Hasan and Memon,  Anis and Rasheed,  Syed Zahed and Memon,  Fazal-ur-Rehman and Mehmood,  Khalid and Ahmed,  Naveeduddin and Qureshi,  Irshad Hussain and Tanveer-us-Salam and Iqbal,  Wasim and Malik,  Uzma and Mehra,  Narinder and Kuo,  Jane Z and Sheu,  Wayne H-H and Guo,  Xiuqing and Hsiung,  Chao A and Juang,  Jyh-Ming J and Taylor,  Kent D and Hung,  Yi-Jen and Lee,  Wen-Jane and Quertermous,  Thomas and Lee,  I-Te and Hsu,  Chih-Cheng and Bottinger,  Erwin P and Ralhan,  Sarju and Teo,  Yik Ying and Wang,  Tzung-Dau and Alam,  Dewan S and Di Angelantonio,  Emanuele and Epstein,  Steve and Nielsen,  Sune F and Nordestgaard,  Børge G and Tybjaerg-Hansen,  Anne and Young,  Robin and Benn,  Marianne and Frikke-Schmidt,  Ruth and Kamstrup,  Pia R and Jukema,  J Wouter and Sattar,  Naveed and Smit,  Roelof and Chung,  Ren-Hua and Liang,  Kae-Woei and Anand,  Sonia and Sanghera,  Dharambir K and Ripatti,  Samuli and Loos,  Ruth J F and Kooner,  Jaspal S and Tai,  E Shyong and Rotter,  Jerome I and Chen,  Yii-Der Ida and Frossard,  Philippe and Maeda,  Shiro and Kadowaki,  Takashi and Reilly,  Muredach and Pare,  Guillaume and Melander,  Olle and Salomaa,  Veikko and Rader,  Daniel J and Danesh,  John and Voight,  Benjamin F and Saleheen,  Danish},
  year = {2017},
  month = sep,
  pages = {1450–1457}
}

@article{James2021,
  title = {The aetiology and molecular landscape of insulin resistance},
  volume = {22},
  ISSN = {1471-0080},
  url = {http://dx.doi.org/10.1038/s41580-021-00390-6},
  DOI = {10.1038/s41580-021-00390-6},
  number = {11},
  journal = {Nature Reviews Molecular Cell Biology},
  publisher = {Springer Science and Business Media LLC},
  author = {James,  David E. and St\"{o}ckli,  Jacqueline and Birnbaum,  Morris J.},
  year = {2021},
  month = jul,
  pages = {751–771}
}

@article{Cox1972,
  title = {Regression Models and Life-Tables},
  volume = {34},
  ISSN = {1467-9868},
  url = {http://dx.doi.org/10.1111/j.2517-6161.1972.tb00899.x},
  DOI = {10.1111/j.2517-6161.1972.tb00899.x},
  number = {2},
  journal = {Journal of the Royal Statistical Society Series B: Statistical Methodology},
  publisher = {Oxford University Press (OUP)},
  author = {Cox,  D. R.},
  year = {1972},
  month = jan,
  pages = {187–202}
}

@article{Bennett2018,
  title = {Inflammation–Nature’s Way to Efficiently Respond to All Types of Challenges: Implications for Understanding and Managing “the Epidemic” of Chronic Diseases},
  volume = {5},
  ISSN = {2296-858X},
  url = {http://dx.doi.org/10.3389/fmed.2018.00316},
  DOI = {10.3389/fmed.2018.00316},
  journal = {Frontiers in Medicine},
  publisher = {Frontiers Media SA},
  author = {Bennett,  Jeanette M. and Reeves,  Glenn and Billman,  George E. and Sturmberg,  Joachim P.},
  year = {2018},
  month = nov 
}

@article{Furman2019,
  title = {Chronic inflammation in the etiology of disease across the life span},
  volume = {25},
  ISSN = {1546-170X},
  url = {http://dx.doi.org/10.1038/s41591-019-0675-0},
  DOI = {10.1038/s41591-019-0675-0},
  number = {12},
  journal = {Nature Medicine},
  publisher = {Springer Science and Business Media LLC},
  author = {Furman,  David and Campisi,  Judith and Verdin,  Eric and Carrera-Bastos,  Pedro and Targ,  Sasha and Franceschi,  Claudio and Ferrucci,  Luigi and Gilroy,  Derek W. and Fasano,  Alessio and Miller,  Gary W. and Miller,  Andrew H. and Mantovani,  Alberto and Weyand,  Cornelia M. and Barzilai,  Nir and Goronzy,  Jorg J. and Rando,  Thomas A. and Effros,  Rita B. and Lucia,  Alejandro and Kleinstreuer,  Nicole and Slavich,  George M.},
  year = {2019},
  month = dec,
  pages = {1822–1832}
}

@article{Vyas2020,
  title = {Hidden in Plain Sight — Reconsidering the Use of Race Correction in Clinical Algorithms},
  volume = {383},
  ISSN = {1533-4406},
  url = {http://dx.doi.org/10.1056/NEJMms2004740},
  DOI = {10.1056/nejmms2004740},
  number = {9},
  journal = {New England Journal of Medicine},
  publisher = {Massachusetts Medical Society},
  author = {Vyas,  Darshali A. and Eisenstein,  Leo G. and Jones,  David S.},
  editor = {Malina,  Debra},
  year = {2020},
  month = aug,
  pages = {874–882}
}

@article{Obermeyer2019,
  title = {Dissecting racial bias in an algorithm used to manage the health of populations},
  volume = {366},
  ISSN = {1095-9203},
  url = {http://dx.doi.org/10.1126/science.aax2342},
  DOI = {10.1126/science.aax2342},
  number = {6464},
  journal = {Science},
  publisher = {American Association for the Advancement of Science (AAAS)},
  author = {Obermeyer,  Ziad and Powers,  Brian and Vogeli,  Christine and Mullainathan,  Sendhil},
  year = {2019},
  month = oct,
  pages = {447–453}
}

@ARTICLE{Nagai2017-ne,
  title     = "Overview of the {BioBank} Japan Project: Study design and
               profile",
  author    = "Nagai, Akiko and Hirata, Makoto and Kamatani, Yoichiro and Muto,
               Kaori and Matsuda, Koichi and Kiyohara, Yutaka and Ninomiya,
               Toshiharu and Tamakoshi, Akiko and Yamagata, Zentaro and
               Mushiroda, Taisei and Murakami, Yoshinori and Yuji, Koichiro and
               Furukawa, Yoichi and Zembutsu, Hitoshi and Tanaka, Toshihiro and
               Ohnishi, Yozo and Nakamura, Yusuke and {BioBank Japan
               Cooperative Hospital Group} and Kubo, Michiaki",
  journal   = "J. Epidemiol.",
  publisher = "Japan Epidemiological Association",
  volume    =  27,
  number    = "3S",
  pages     = "S2--S8",
  month     =  mar,
  year      =  2017,
  copyright = "http://creativecommons.org/licenses/by-nc-nd/4.0/",
  language  = "en"
}

@article{Brown2020,
  title = {Using and improving distributed data networks to generate actionable evidence: the case of real-world outcomes in the Food and Drug Administration’s Sentinel system},
  volume = {27},
  ISSN = {1527-974X},
  url = {http://dx.doi.org/10.1093/jamia/ocaa028},
  DOI = {10.1093/jamia/ocaa028},
  number = {5},
  journal = {Journal of the American Medical Informatics Association},
  publisher = {Oxford University Press (OUP)},
  author = {Brown,  Jeffrey S and Maro,  Judith C and Nguyen,  Michael and Ball,  Robert},
  year = {2020},
  month = apr,
  pages = {793–797}
}

@ARTICLE{Caruana1997-es,
  title     = "Multitask learning",
  author    = "Caruana, Rich",
  journal   = "Mach. Learn.",
  publisher = "Springer Science and Business Media LLC",
  volume    =  28,
  number    =  1,
  pages     = "41--75",
  month     =  jul,
  year      =  1997,
  copyright = "https://www.springernature.com/gp/researchers/text-and-data-mining",
  language  = "en"
}

@misc{https://doi.org/10.13026/hxp0-hg59,
  doi = {10.13026/HXP0-HG59},
  url = {https://physionet.org/content/mimiciv/3.0/},
  author = {Johnson,  Alistair and Bulgarelli,  Lucas and Pollard,  Tom and Gow,  Brian and Moody,  Benjamin and Horng,  Steven and Celi,  Leo Anthony and Mark,  Roger},
  title = {MIMIC-IV},
  publisher = {PhysioNet},
  year = {2024}
}

@article{Sudlow2015,
  title = {UK Biobank: An Open Access Resource for Identifying the Causes of a Wide Range of Complex Diseases of Middle and Old Age},
  volume = {12},
  ISSN = {1549-1676},
  url = {http://dx.doi.org/10.1371/journal.pmed.1001779},
  DOI = {10.1371/journal.pmed.1001779},
  number = {3},
  journal = {PLOS Medicine},
  publisher = {Public Library of Science (PLoS)},
  author = {Sudlow,  Cathie and Gallacher,  John and Allen,  Naomi and Beral,  Valerie and Burton,  Paul and Danesh,  John and Downey,  Paul and Elliott,  Paul and Green,  Jane and Landray,  Martin and Liu,  Bette and Matthews,  Paul and Ong,  Giok and Pell,  Jill and Silman,  Alan and Young,  Alan and Sprosen,  Tim and Peakman,  Tim and Collins,  Rory},
  year = {2015},
  month = mar,
  pages = {e1001779}
}

@article{Collins2024TRIPODAI,
  title = {TRIPOD+AI statement: updated guidance for reporting clinical prediction models that use regression or machine learning methods},
  author = {Collins, Gary S and Moons, Karel G M and Dhiman, Paula and Riley, Richard D and Beam, Andrew L and Van Calster, Ben and Ghassemi, Marzyeh and Liu, Xiaoxuan and Reitsma, Johannes B and van Smeden, Maarten and Boulesteix, Anne-Laure and Camaradou, Jenny and Celi, Leo Anthony and Denaxas, Spiros and Denniston, Alastair K and Glocker, Ben and Golub, Robert M and Harvey, Hugh and Heinze, Georg and Hoffman, Michaela M and Kengne, Andre Pascal and Lam, Cecilia and van der Schaar, Mihaela and Vollmer, Sebastian J and Wilkinson, Jack and Yang, Lin and Lones, Michael A},
  journal = {BMJ},
  volume = {385},
  pages = {e078378},
  year = {2024},
  doi = {10.1136/bmj-2023-078378}
}

@article{vonElm2007STROBE,
  title = {The Strengthening the Reporting of Observational Studies in Epidemiology (STROBE) statement: guidelines for reporting observational studies},
  author = {von Elm, Erik and Altman, Douglas G and Egger, Matthias and Pocock, Stuart J and G{\o}tzsche, Peter C and Vandenbroucke, Jan P},
  journal = {PLOS Medicine},
  volume = {4},
  number = {10},
  pages = {e296},
  year = {2007},
  doi = {10.1371/journal.pmed.0040296}
}

\clearpage
%\onecolumn
%\newpage

% \end{appendix}
\end{document}